\documentstyle[psfig]{l-aa}

\begin{document}

   \thesaurus{1         
              (12.03.3;  
09.01.1; 
10.01.1; 
13.19.3; 
               )} 
   \title{ DI in the outer Galaxy }
   \author{ Jayaram N. Chengalur \inst{1}
     \and   Robert Braun \inst{1}
     \and   W. Butler Burton \inst{2}
          }

   \offprints{Robert Braun}

   \institute{Netherlands Foundation for Research in Astronomy, P.O. Box 2,
              7990 AA Dwingeloo,
             The Netherlands
    \and     Sterrewacht Leiden,
             P.O. Box 9513, 2300 RA Leiden,
             The Netherlands
             }

   \date{Received ----; accepted -------}

   \maketitle

   \begin{abstract}
     
     We report on a deep search with the Westerbork Synthesis Radio
     Telescope towards the galactic anticenter for the 327~MHz
     hyperfine transition of DI. This is a favorable direction for a
     search because:~(i)~the HI optical depth is high due to
     velocity crowding; (ii)~the observed molecular column density is
     low (implying that most of the deuterium would probably be in atomic
     form, rather than in HD); and (iii)~the stellar reprocessing
     should be minimal.
     
     Our observations are about a factor of two more sensitive than
     previous searches for DI in this direction. We detect a low
     significance ($\sim4~\sigma$) feature, consistent in both
     amplitude and center frequency with an emission feature reported
     previously (Blitz \& Heiles 1987).  If this is the DI line, then
     the implied $N_{\rm D}/N_{\rm H}$ of $3.9\pm1.0\times10^{-5}$
     is comparable to the inferred pre-solar deuterium abundance. Our
     observation is consistent with the recent low measurements of D/H
     towards high-redshift Lyman-limit systems. On the other hand, if
     the reports of high DI abundance ($\sim 24\times10^{-5}$) in such
     systems are confirmed, then our observations imply that even in
     regions of reduced star formation within the outer Galaxy, the DI
     abundance has been reduced by a factor of $\sim6$ from the
     primordial abundance.


   \end{abstract}

%

\section{Introduction}
        
In the standard big bang model, the primordial abundance
of deuterium is a sensitive function of the baryon to photon ratio,
(e.g. Walker et al. 1991) making it a quantity of great cosmological
interest. Further, since all known astrophysical processes (apart from
the big bang itself, of course) result in a net destruction of
deuterium, the currently observed value of the deuterium abundance is
a strict lower limit to its primordial abundance.

HST observations of the DI Lyman$-\alpha$ line in the local solar
neighborhood (Linsky et al. 1993) yield a deuterium abundance of
$N_{\rm D}/N_{\rm H} = 1.65^{+0.07}_{-0.18}\times 10^{-5}$. Conversion
from this local current abundance to the primordial abundance depends
on less well understood details of the history of the stellar
reprocessing of matter in the local ISM. In order to circumvent this
problem a number of groups have been attempting to measure the
deuterium abundance in high-redshift Lyman-limit systems.  Since these
systems are less chemically evolved than the local ISM, conversion
from the measured to the primordial abundance should be more
straightforward. However, the results of these observations are
conflicting, with different groups (Carswell et al. 1994, Songalia et
al. 1994, Tytler et al. 1996) measuring abundances which differ by
more than an order of magnitude.  Part of the problem is that the DI
Lyman$-\alpha$ line is separated from the HI Lyman$-\alpha$ line by
only 82~km~s$^{-1}$, making the chance of contamination of the DI line
by absorption from a small parcel of HI at a slightly different
velocity from that of the main Lyman-limit system non-negligible.

There have also been several attempts to observe the hyperfine
transition of DI at radio frequencies. Since the frequency of this
line is more than a factor of 4 lower than the frequency of the
corresponding transition in HI, the question of HI contamination does
not arise. Two kinds of lines-of-sight have been favored in the past,
the first towards bright radio sources (Sgr~A \& Cas~A; Weinreb 1962,
Anantharamiah \& Radhakrishnan 1979, Heiles et al. 1993), where the
hydrogen column density is known to be high. The disadvantages of
these lines of sight are that: (i)~the bright radio sources contribute
significantly to the system temperature, making detection more
difficult; (ii)~any measurement refers only to the thin pencil beam
subtended by the absorber; and (iii)~the molecular column density is
also high, making it likely that most of the deuterium is in molecular
rather than atomic form (Heiles et al. 1993). Anantharamiah \&
Radhakrishnan (1979) placed an upper limit of $5.8\times 10^{-5}$ on
the DI abundance towards Sgr~A. Heiles et al. (1993) reached similar
limits towards Sgr~A as well as Cas~A.

The other promising direction for a search for the radio emission from
DI is that towards the galactic anticenter, where one expects the line
to be in emission. The advantages of this direction are that (i)~the high
optical depth of HI is due to velocity crowding along a long
pathlength rather than a high volume density; (ii)~the molecular column
density and metallicity are low; and (iii)~the observations are
sensitive to the DI abundance within the entire telescope beam, and
not just a narrow cone towards the background source as in the case of
absorption observations.

The results of a long integration in the direction $(l,b)~=~(183\fdg
0,+0\fdg 5)$ using the Hat Creek telescope were presented by Blitz \&
Heiles (1987), who found an upper limit ($2~\sigma$) of $6.0\times
10^{-5}$ for $N_{\rm D}/N_{\rm H}$.  Here we report on a long
integration towards a partially overlapping line-of-sight,
$(l,b)~=~(183\fdg 0,-0\fdg 5)$, with the Westerbork synthesis array.
Using the 14 WSRT telescopes as independent single dishes allowed us to
significantly increase the effective integration time.

%

\section{ Observations and data reduction }

\label{sec:obs}

The observations were conducted at the WSRT during 29 sessions between
9~March and 14~May 1989, and used a special correlator configuration
which provided 28 auto-correlations (14~telescopes $\times$
2~polarizations) in addition to a number of cross-correlations between
the different telescopes.  The cross-correlation data were not used in
the data analysis.  The observations were done in a frequency switched
mode, with the magnitude of the frequency throw equal to the total
bandwidth of 0.156 MHz ($\sim 140$~km~s$^{-1}$). Each spectrum had 256
uniformly tapered channels. The ON spectrum was centered on $V_{\rm
  LSR} = 0$~km~s$^{-1}$.  Either before or after each source
observation (which typically lasted a few hours) a standard calibrator
was observed in the same frequency switched mode for about an hour.

The data reduction was done in WASP (Chengalur 1996), a package
suited to the automated reduction of large quantities of spectral data.
For each observation, the data for each polarization of each telescope
were reduced separately. Every 60 second ON spectrum was baselined using
the corresponding  OFF spectrum. Next the following flagging and
baselining operations were done iteratively, until the the rms of the
remaining unflagged data converged (which it typically did within 3
iterations). The rms over all channels and all time was computed, all
points which exceeded 6 sigma were flagged. A linear baseline was then
fit and subtracted from  each spectrum (excluding these flagged
points). Spectra whose rms (over channel numbers) exceeded 2.5 times the
global rms were flagged, and  finally channels whose rms (over time)
exceeded 3 times the global rms were flagged. For all but the first
iteration, the previously fit baseline was first added back to each
spectrum. These criteria were selected after trial runs on a few sample
observations and were tailored for efficient removal of narrow-band EMI
and of spectra in which the EMI was strong enough to cause ringing
effects. The data from telescopes 4, 5, 6 and 7 (which are the closest
to the control building) showed a large amount of EMI, and are almost
entirely flagged out. Finally, all the unflagged data were averaged
together to produce one average spectrum per polarization per telescope
per observation.

Since low-level intermittent EMI sometimes survived this flagging
process, this average spectrum was inspected visually and telescopes
with broad multi-channel EMI or with low-level ringing were flagged.
All the remaining data were averaged together, separately for each
polarization of each telescope. These processing steps were carried
out separately for the source and the calibrator observations. A
4$^{\rm th}$
order polynomial baseline was then fit to the {\it calibrator }
spectrum and then this {\it same} baseline was removed from the source
observations (after allowing for a small scaling difference between
the source and calibrator observation). The spectra from two telescopes 
(both polarizations of RTD and one polarization of RT3) showed substantial
baseline structure even after this correction and were dropped from
further processing. The spectra from the remaining telescopes were
then averaged together and smoothed with a Gaussian smoothing function
of 11.2~km~s$^{-1}$ FWHM. A few channels which still showed unresolved
peaks,  presumably due to EMI, were blanked before
smoothing.  Since the smoothing length was $\sim 20$ channels, this had
very little effect on the spectrum. 

The final spectrum (excluding about 10\% of the band on either edge
for which the noise level is higher and the calibrator baseline is
more poorly determined) is shown in Figure~1.  The peak-to-peak
variation is about 4~mK, approximately a factor of two smaller than
that of the spectrum in Blitz \& Heiles (1987). It is interesting to
note that similar to Blitz \& Heiles, there is a feature of low
significance ($\sim 3~\sigma$ peak brightness) at the velocity $\sim
7$~km~s$^{-1}$.  We have overlaid a Gaussian profile with peak
brightness 2.4~mK and the same velocity centroid, $+7$~km~s$^{-1}$ and
FWHM of 18.7~km~s$^{-1}$ as the HI emission feature in this direction.
Excluding this possible feature, the noise level is 0.85~mK over
11.2~km~s$^{-1}$, again about a factor of 2 lower than the 1.6~mK
quoted in Blitz \& Heiles 1987 (who also appear to have excluded the
putative feature before computing the noise level).  Including this
feature in the calculation of the noise gives an rms of 1.06~mK, while
the expected thermal noise is 0.8~mK.

\begin{figure*}[htbp]

\centerline{
\psfig{file=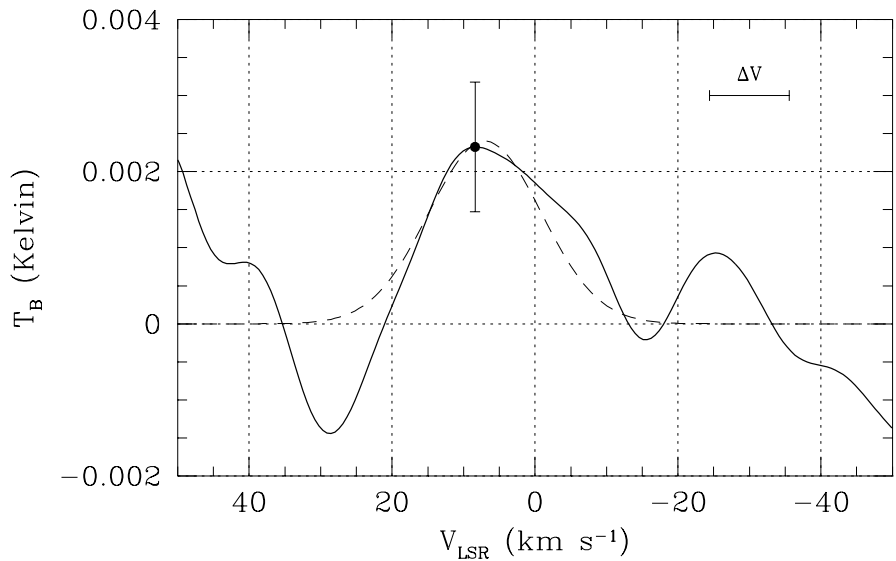,height=8cm,bbllx=65pt,bblly=234pt,bburx=576pt,bbury=558pt,clip=}
}

\caption{Emission spectrum at the frequency of the DI $\lambda$92-cm 
  hyperfine transition in the direction $(l,b)~=~(183\fdg 0,-0\fdg 5)$. The
  rms fluctuation level, excluding the possible feature near
  7~km~s$^{-1}$, is 0.85~mK and is consistent with the thermal noise.
  The velocity resolution, of 11.2~km~s$^{-1}$, is indicated at the
  upper right. A Gaussian profile with peak amplitude 2.4~mK and the
  same velocity centroid, V$_{\rm LSR}~=~+7$~km~s$^{-1}$ and FWHM of
  18.7~km~s$^{-1}$ as the HI emission feature in this direction is
  overlaid.
}
\label{Fig1}
\end{figure*}

\section{ Discussion }

The galactic anticenter region, near $(l,b)~=~(183\fdg 0,0\degr )$, is
particularly well-suited to a search for the DI hyperfine emission
feature at $\lambda$92-cm. The HI emission brightness in this region
peaks smoothly into a plateau extending spatially over several degrees
with a brightness temperature 135$\pm$3~K, as can be seen in Burton
(1985).  Conditions should therefore be reasonably constant over the
$2\fdg 5$ beam of the WSRT dishes.  The extended region of line
brightening is paired with a line narrowing due to velocity crowding.
The HI emission-line profile from this region is relatively narrow,
with a FWHM of 18.7~km~s$^{-1}$. The line core is rather flat-topped,
strongly suggesting a high line opacity. It also displays shallow
self-absorption, indicating a modest amount of temperature
substructure along the line-of-sight.  Based on the depth of the
self-absorption features and assuming a substantial line opacity, the
spin temperature of the atomic gas appears to be in the range
125--135~K.

Direct measurements of the HI opacity along many lines of sight in
this region have not yet been carried out. The typical sky density of
moderately bright extragalactic background sources suitable for such a
measurement suggests that a few tens of lines of sight could, in
principle, be observed. The closest line of sight that has been
observed in HI absorption (Dickey et al. 1983) is at $(l,b)~=~(189\fdg
6,-0\fdg 6)$. This is near the edge of the plateau in emission
brightness, where the line profile is already somewhat broader than in
the direction of maximum velocity crowding, with a FWHM of
23.4~km~s$^{-1}$.  The absorption profile extends smoothly over the
entire velocity extent in which the emission profile exceeds a
brightness of about 5--10~K, with a uniform high opacity of about 2
across the line core. This is consistent with the expectation for a
gas distribution in which the column density is dominated by an
approximately isothermal gas with kinetic temperature in the range
125--135~K. An absorption equivalent width, 
$\int\tau~{\rm d}V$~=~33.68~km~s$^{-1}$ is found. 
The very smooth nature of the HI
emission suggests that it is plausible to expect a comparable
equivalent width in the direction
$(l,b)~=~(183\fdg 0,-0\fdg 5)$.

Assuming that the DI emission is either mixed with, or lies behind, most
of the galactic continuum emission, (which has the substantial
brightness of about 70~K at 92~cm wavelength) the equation of
radiative transfer yields simply:

\begin{equation}
T_{\rm B} = T_{\rm S}(1- e^{-\tau_{\rm D}})
\end{equation}
        
where $T_{\rm B}$ is the differential brightness temperature on and
off the line, $T_{\rm S}$ is the spin temperature of the DI and
${\tau_{\rm D}}$ is the optical depth of the DI. 
For ${\tau_{\rm D}}~<<~1$ this yields 
$T_{\rm B}~=~T_{\rm S}\tau_{\rm D}$, or
$\int~T_{\rm B}~{\rm d}V~=~T_{\rm S}\int\tau_{\rm D} {\rm d}V$ for an
approximately isothermal gas. The line integral of the Gaussian
overlaid on the possible feature in Figure~1 is 
$\int~T_{\rm B}~{\rm d}V~=~0.048\pm0.012$~K~km~s$^{-1}$.  
Assuming that the spin
temperature of the DI is the same as that of HI (i.e. 130~K), this
corresponds to 
$\int\tau_{\rm D}~{\rm d}V$~=~3.7$\pm0.9\times$10$^{-4}$~km~s$^{-1}$.
Hence the estimate
of the ratio of the optical depths is 
${\tau_{\rm D}/\tau_{\rm H}}~=~1.1\pm0.3 \times 10^{-5}$. 
The relationship between the
ratio of the optical depths and the ratio of the column densities of
DI and HI is (Anantharamiah \& Radhakrishnan 1979)
${\tau_{\rm D}/\tau_{\rm H}}~=~0.28~{N_{\rm D}/N_{\rm H}}$, hence the estimated
abundance of DI is $3.9\pm1.0\times10^{-5}$.  This is comparable to
the inferred pre-solar abundance (Kunde et al.  1982, Courtin et al.
1984), and about a factor of 2 above the current DI abundance in the
local solar neighborhood (Linsky et al. 1993).

As discussed in the introduction, the lack of detection of DI towards
the inner Galaxy suggests that most of the DI has been converted
into HD within molecular clouds (Heiles et al. 1993). Towards the
outer Galaxy, however, the molecular column density is low along many
sight lines. In fact, the survey by Dame et al. (1987) detected very
little CO emission in the galactic plane between 180 and 185 degrees
longitude except for a small concentration at $-10$~km~s$^{-1}$.  The
high optical depth in HI seems to be due to velocity crowding in
diffuse gas along a pathlength of several kpc, and does not appear to
arise in a single physical entity like a giant molecular cloud. It
seems plausible, therefore, that a large fraction of the deuterium in
this direction is still in atomic form. Further, the metallicity
gradient in the Galaxy suggests that the DI abundance in the outer
Galaxy would be higher than the inner. Calculations by Prantzos (1996)
show that the deuterium abundance gradient is in general steeper than
(and of course opposite in sign to) the oxygen gradient (because of the
late ejection of metal-poor but deuterium-free gas from low mass stars
formed at early times). While the exact gradient is sensitive to the
assumed infall model, the current DI abundance within diffuse atomic
gas in the inner and outer Galaxy could well differ by a factor of
two.

Another useful comparison to make is with the range of measured DI
abundances, $2.3-24\times10^{-5}$, in high redshift Lyman limit
systems (Carswell et al. 1994, Songalia et al. 1994, Tytler et al.
1996). If a high primordial abundance turns out to be correct, our
results indicate that even in regions with reduced star formation, the
current deuterium abundance is about a factor of 6 lower than
primordial. In contrast, recent model calculations of astration of
deuterium suggest that the abundance evolution is modest, with current
abundance only a factor $\sim2$ less than primordial (Galli et al.
1995).  On the other hand, the low value of D/H measured by Tytler et
al. (1996), $2.3\pm0.3\times10^{-5}$, is consistent with our own
possible detection and the inferred pre-solar value. Direct imaging of
DI emission in the outer regions of nearby galaxies may provide a very
effective means of addressing this issue comprehensively, once
the capability for achieving sub-mK sensitivity on arcmin scales
becomes available with the next generation of radio telescopes. The
Giant Meterwave Radio Telescope (GMRT) should already allow a robust
detection of DI emission from the Galaxy within the next few years,
while DI imaging of nearby external galaxies should become possible
with construction of the Square Kilometer Array Interferometer (SKAI)
early in the mext millenium.

\begin{acknowledgements}
  
  We would like to thank Hans van Someren Gr\'eve and the WSRT staff
  for carrying out these non-standard observations, and Bauke Kramer,
  Tony Foley and Geert Kuiper for assistance in retrieving the data
  from the WSRT archive.

\end{acknowledgements}


\begin{thebibliography}{}

   \bibitem{} Anantharamiah, K. R. \& Radhakrishnan, V., 1979, 
        A\&A 79, L9

   \bibitem{} Blitz, L. \& Heiles, C., 1987, ApJ 313, L95

\bibitem{} Burton, W.B., 1985, A\&AS 62, 365

   \bibitem{}Carswell, R. F., Rauch, M., Weyman, R. J., Cooke, A. J.
        \& Webb, J. K., 1994, MNRAS 268, L1 

   \bibitem{}Chengalur, J. N., 1996, NFRA Note 453

   \bibitem{}Courtin, R., Gautier, D., Marten, A. \& Bezard, B., 1984,
        ApJ 287, 899       

   \bibitem{} Dame, T. M. et al., 1987, ApJ 322, 706

\bibitem{} Dickey, J.M., Kulkarni, S.R., van Gorkom, J.H. \& Heiles,
  C.E., 1983, ApJS 53, 591 

   \bibitem{} Heiles, C., McCullogh, P. R., \& Glassgold, A. E., 1993,
        ApJS 89, 271

   \bibitem{} Galli, D., Palla, F., Ferrini, F. and Penco, U., 1995
        ApJ 443, 536

   \bibitem{} Kunde, V.G. et al., 1982, ApJ 263, 443        

   \bibitem{} Linsky, J. L. et al., 1993, ApJ 402, 694

   \bibitem{} Prantzos, N., 1996, A\&A 310, 106

   \bibitem{} Songalia, A., Cowie, L. L., Hogan, C. J. \& Rugers. M., 1994
        Nat 368, 599

\bibitem{} Tytler, D., Fan, X.-M. \& Burles, S., 1996, Nat 381, 207

   \bibitem{} Walker, T. P., Steigman, G., Schramm, D. N., Olive, K. A.
        \& Kang, H. S., 1991, ApJ 376, 51 

   \bibitem{} Weinreb. S., 1962, Nat 195, 367



\end{thebibliography}
\end{document}